\documentclass[letterpaper,twocolumn,english,prl,superscriptaddress]{revtex4-1}
\usepackage[latin9]{inputenc}
\usepackage{amsmath}
\usepackage{amssymb}
\usepackage{graphicx}



\makeatother

\usepackage{babel}
\begin{document}

\title{Tunable Intrinsic Plasmons due to Band Inversion in Topological Materials}

\author{Furu Zhang}

\affiliation{Beijing Key Laboratory of Nanophotonics and Ultrafine Optoelectronic
Systems, School of Physics, Beijing Institute of Technology, Beijing
100081, China}

\author{Jianhui Zhou}
\email{jianhuizhou1@gmail.com}

\affiliation{Department of Physics, Carnegie Mellon University, Pittsburgh, Pennsylvania
15213, USA}

\affiliation{Department of Physics, The University of Hong Kong, Pokfulam Road,
Hong Kong, China}

\author{Di Xiao}

\affiliation{Department of Physics, Carnegie Mellon University, Pittsburgh, Pennsylvania
15213, USA}

\author{Yugui Yao}
\email{ygyao@bit.edu.cn}

\selectlanguage{english}%

\affiliation{Beijing Key Laboratory of Nanophotonics and Ultrafine Optoelectronic
Systems, School of Physics, Beijing Institute of Technology, Beijing
100081, China}

\begin{abstract}
The band inversion has led to rich physical effects in both topological
insulators and topological semimetals. It has been found that the inverted band structure with the Mexican-hat dispersion could enhance the interband correlation
leading to a strong intrinsic plasmon excitation. Its frequency ranges
from several $\mathrm{meV}$ to tens of $\mathrm{meV}$ and can be
effectively tuned by the external fields. The electron-hole asymmetric
term splits the peak of the plasmon excitation into double peaks.
The fate and properties of this plasmon excitation can also act as
a probe to characterize the topological phases even in the lightly
doped systems. We numerically demonstrate the impact of the band inversion
on plasmon excitations in magnetically doped thin films of three-dimensional
strong topological insulators, V- or Cr-doped (Bi,Sb)$_2$Te$_3$, which support the quantum anomalous
Hall states. Our work thus sheds some new light on the potential applications of topological materials in plasmonics. 
\end{abstract}


\maketitle
\textit{Introduction.}\textendash Plasmons are ubiquitous collective
density oscillations of an electron liquid and can occur in metals,
doped semiconductors and semimetals \cite{maier2007plasmonics,giuliani2005qtel}.
The frequency of plasmon excitation is usually proportional to the
density of states (DOS) of carriers at the Fermi level in the long
wavelength limit \cite{pines1999elementary}. In systems
with vanishing DOS at the Fermi level, however, it had been demonstrated
that strong correlation, higher order term in the effective mass, and
the chiral anomaly could give rise to exotic plasmon excitations 
at zero temperature in the context of two dimensional (2D) Anderson
insulators \cite{Shahbazyan1996PRB}, graphene \cite{Gangadharaiah2008PRL},
HgTe quantum well \cite{Juergens2014PRL}, and three-dimensional (3D) Weyl
semimetals \cite{Zhou2015Plasmon}, respectively. In this paper, we
reveal a new mechanism based on the band inversion with the Mexican-hat dispersion 
that can be used to excite and manipulate intrinsic plasmon excitations in topological materials even with vanishing DOS. 

The band inversion has played a critical role in the transport properties
of both topological insulators and topological semimetals, which have
attracted intensive interest in condensed matter physics~\cite{Hasan2010RMP,QiZhangRMP,Bansil2016RMP,Armitage2017rmp}.
However, there has been little study of the effect of band inversion on the optical
properties of topological materials~\cite{Tan2016PRL}.
Recently, the experimental observation of the quantum anomalous Hall
(QAH) effect \cite{HaldaneQAHE1988,Yu2010Science} has been reported
in magnetically doped ultrathin films of topological insulators V- or Cr-doped (Bi,~Sb)$_2$Te$_3$~\cite{chang2013QAHEScience,Bestwick2015QAHE,chang2015QAHE},
in which its band topology can be tuned by external fields. On the
other hand, some recent experimental efforts have been made on observing
the plasmon excitations in the bulk or surface of both the magnetic
and nonmagnetic 3D topological insulators \cite{Raghu2010PRL,Profumo2012PRB,diPietro2013,ou2014plasmonTI,Politano2015PRL,Kogar2015PRL,Glinka2016NC,mahoney2017EDGE,Jia2017PRL}.
The frequency of these plasmon excitations ranges from terahertz
to far infrared and facilitates a wide variety of applications
of topological insulators, such as in information and communication,
chemical and biological sensing, and medical sciences \cite{Ferguson2002NatM,Tonouchi2007NPhot,Soref2010NPhot}.
Therefore, these QAH topological materials provide us with a promising
platform to controllably create and manipulate collective excitations
via the band inversion mechanism, a new control knob for the field of
plasmonics~\cite{maier2007plasmonics}.  

In this paper, we show that the inverted band structure with the Mexican-hat dispersion
could enhance the interband correlation, leading to a strong intrinsic
plasmon excitation.  We theoretically examine
the relevant physics in both the intrinsic and lightly doped thin films of topological insulators. 
The frequency of the resulting intrinsic plasmon
can be effectively tuned by the external fields. The peak of plasmon
excitation is split into double peaks by the electron-hole asymmetry.
Consequently, the plasmon excitation provides us with an effective
tool to identify the topological phases. %

\textit{Model of the QAH systems.}\textendash To demonstrate our theory, 
we begin with the following Hamiltonian that describes the band inversion 
in the magnetically doped thin films of topological insulators V- or Cr-doped (Bi,~Sb)$_2$Te$_3$~\cite{LuHZ2013PRL} 
\begin{equation}
H=H_{0}+\frac{m}{2}\tau_{0}\otimes\sigma_{z},\label{Ham}
\end{equation}
where $m$ is the exchange field originating from the magnetic dopants,
effectively acting as a Zeeman field. $H_{0}$ is given by~\cite{zhang2009TI,Shan2010NJP}
\begin{equation}
H_{0}=-Dk^{2}+\left(\begin{array}{cc}
h_{+}\left(\boldsymbol{k}\right) & V\\
V & h_{-}\left(\boldsymbol{k}\right)
\end{array}\right),\label{H0}
\end{equation}
with 
\[
h_{\pm}\left(\boldsymbol{k}\right)=\left(\begin{array}{cc}
\pm(\frac{\Delta}{2}-Bk^{2}) & iv_{F}k_{-}\\
-iv_{F}k_{+} & \mp(\frac{\Delta}{2}-Bk^{2})
\end{array}\right),
\]
where $h_{\pm}\left(\boldsymbol{k}\right)$ describes the 2D Dirac
fermions with a $k$-dependent mass. Here $\boldsymbol{k}=(k_{x},k_{y})$
is the 2D wave vector, $k_{\pm}=k_{x}\pm ik_{y}$ and $k=\sqrt{k_x^2+k_y^2}$, $v_{F}$ is
the effective velocity, the $D$ term breaks the electron-hole symmetry,  
$\Delta$ is the hybridization of the top and bottom surface states
of the thin film, and $V$ measures the structural inversion asymmetry
between the top and bottom surfaces. 
The Pauli matrices $\tau_{0}$ and $\sigma_{z}$
act on the pseudospin space related to the top and bottom surfaces
and the real spin degree of freedom, respectively.  In this paper we primarily
consider the insulating phase that requires $\left|D\right|<\left|B\right|$.
The actual value of the parameters describes realistic materials~\cite{Yu2010Science,chang2013QAHEScience}.
Specifically, $v_{F}=3.0\:\mathrm{eV}\cdot\mathbf{\mathrm{\mathring{A}}}$,
$V=0.03\:\mathrm{eV}$ and $B=-30\:\mathrm{eV}\cdot\mathbf{\mathrm{\mathring{A}}}^{2}$. 

\begin{figure}
\centering{}\includegraphics[scale=0.95]{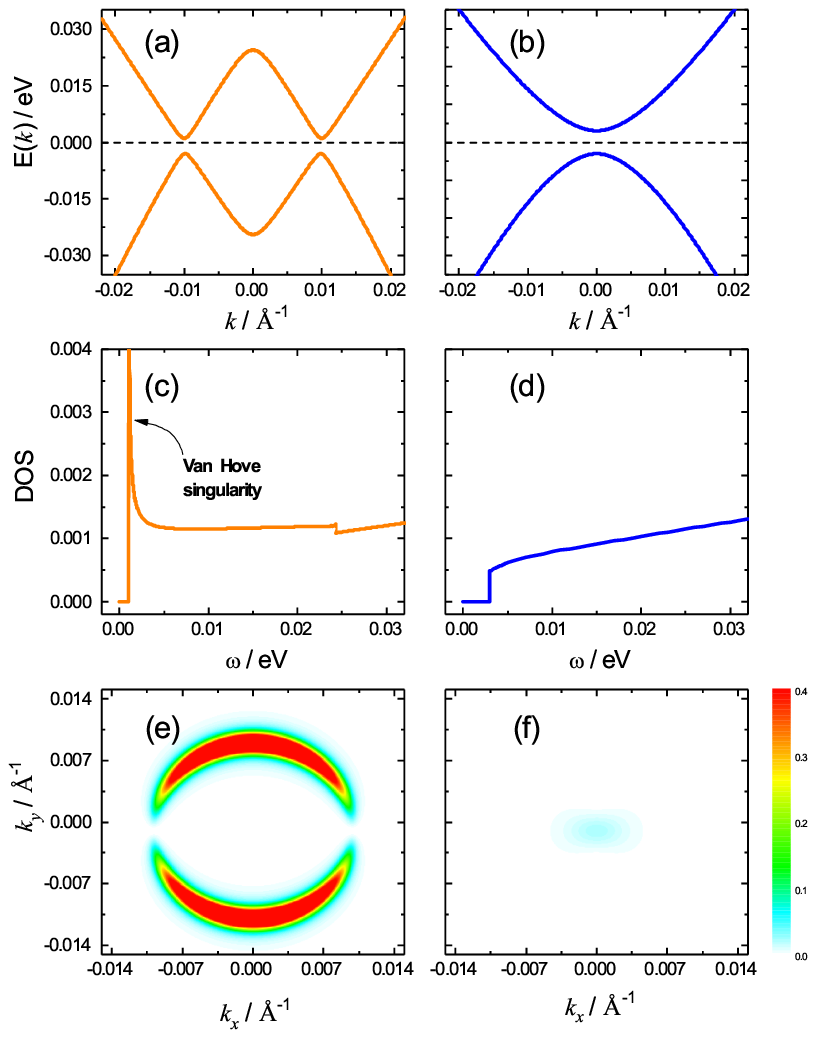}\caption{(color online) The band structures of the two inner bands of the thin film of topological
insulators described by Eq. $\left(\ref{Ham}\right)$ in the trivial (a) and nontrivial (b) phases, respectively. 
(c) and (d) show the corresponding DOS. (e) and (f) are the intensity plots  
of the overlap of wave functions $\mathit{F_{\mathrm{2,3}}\mathrm{(}\boldsymbol{k},\boldsymbol{k}^{\prime}\mathrm{)}}$. 
Parameters: $\Delta=-0.01\:\mathrm{eV}$, $D=10.0\:\mathrm{eV}\cdot\mathbf{\mathrm{\mathring{A}}}^{2}$,
$\eta=10^{-5}$ eV and $q=2\times10^{-3}\:\mathbf{\mathrm{\mathring{A}{}^{-1}}}$.
\label{FigBDO}}
\end{figure}

Let us first discuss the topological properties of the Hamiltonian in Eq. $\left(\ref{Ham}\right)$. As the exchanged
field $m$ increases, the system will undergo a topological phase
transition from a trivial phase to a nontrivial one or vice versa. 
The critical value of the exchanged filed is $m_{0}=-\sqrt{\Delta^{2}+4V^{2}}$.
It is useful to introduce a dimensionless parameter $\xi=m/m_{0}$
to mark the topological phase transition. When $\xi<1$, the system
is in a topologically trivial phase, and when $\xi>1$, it is in a
nontrivial phase with chiral edge states~\cite{chang2013QAHEScience}.
The band gap between the two inner bands closes at the critical point
with $\xi=1$. Fig. \ref{FigBDO}(a) clearly exhibits that in the trivial
phase, the band edge have an obvious Mexican-hat like
inverted band structure. The joint density of states near the band
edge of this inverted band exhibits a 1D behavior and possesses a Van
Hove singularity at a finite wave vector, in which the corresponding
DOS diverges, as shown in Fig. \ref{FigBDO}(c). On the other hand,
in the nontrivial phase [see Fig. \ref{FigBDO}(b)], all the four bands are parabolic, i.e., uninverted
\cite{LuHZ2013PRL}. The DOS near these band edges does not possess
a strong singularity [Fig.~\ref{FigBDO}(d)]. 
In addition, Figs. \ref{FigBDO}(e)-\ref{FigBDO}(f) show a great enhancement of the overlap of wave functions due to the band inversion.  

\textit{Band inversion mechanism for plasmons.}\textendash We start
from the general dielectric function at arbitrary wave vector $\boldsymbol{q}$
and frequency $\omega$ within the random phase approximation  \cite{pines1999elementary}
\begin{equation}
\varepsilon(\boldsymbol{q},\omega)=1-V(q)\Pi(\boldsymbol{q},\omega),\label{dief}
\end{equation}
where $\mathit{V\mathrm{(\mathit{q})=2\pi\mathit{e^{\mathrm{2}}/\kappa q}}}$
is the Fourier transform of the 2D Coulomb interaction, $\kappa$
is the effective background dielectric constant. The noninteracting
polarization function $\Pi(\boldsymbol{q},\omega)$ has the form~\cite{SMeq4} 
{\small{}
\begin{equation}
\Pi(\boldsymbol{q},\omega)=\underset{\lambda,\lambda^{\prime}}{\sum}\int\frac{d^{2}\boldsymbol{k}}{(2\pi)^{2}}\frac{f(E_{\lambda k})-f(E_{\lambda^{\prime}k^{\prime}})}{\omega+E_{\lambda k}-E_{\lambda^{\prime}k^{\prime}}+i\eta}F_{\lambda\lambda^{\prime}}(\mathit{\boldsymbol{k},\boldsymbol{k}^{\prime}}),\label{polf}
\end{equation}
}in which the overlap of eigenstates $\mathit{F_{\lambda\lambda^{\prime}}\mathrm{(}\boldsymbol{k},\boldsymbol{k}^{\prime}\mathrm{)}}$
is given by
\begin{equation}
F_{\lambda\lambda^{\prime}}\left(\mathit{\boldsymbol{k},\boldsymbol{k}^{\prime}}\right)=\left|\left\langle \boldsymbol{k},\lambda|\lambda^{\prime},\boldsymbol{k}^{\prime}\right\rangle \right|^{2},\label{overlp}
\end{equation}
where $\left|\lambda^{\prime},\boldsymbol{k}^{\prime}\right\rangle $
is the periodic part of the Bloch wave function with $\boldsymbol{k}^{\prime}=\boldsymbol{k}+\boldsymbol{q}$.
The band indices $\lambda$ and $\lambda^{\prime}$ run over all relevant
energy bands. $\eta$ is related to the electron lifetime due to the
Landau damping. $\mathit{E_{\lambda k}}$ is the energy
dispersion of the effective Hamiltonian, and $f\left(x\right)=\left[\exp\left\{ \beta\left(x-\mu\right)\right\} +1\right]^{-1}$
is the Fermi distribution function with $\beta=1/k_{B}T$. For simplicity,
we assume zero temperature $T=0\:\mathrm{K}$ throughout this paper.
It is well known that the plasmons can be revealed as sharp peaks
in the energy-loss function, defined as the imaginary part of the
inverse dielectric function \cite{pines1999elementary}
\begin{equation}
\mathrm{Loss\left(\mathit{\boldsymbol{q}},\mathit{\omega}\right)}=\mathrm{Im}\left[-\frac{1}{\varepsilon\left(\mathit{\boldsymbol{q}},\mathit{\omega}\right)}\right],
\end{equation}
which can be probed in various spectroscopy experiments \cite{grigorenko2012graphene},
such as the electron energy-loss spectroscopy and angle-resolved photoemission spectroscopy. 
\begin{figure}
\begin{centering}
\includegraphics[scale=0.65]{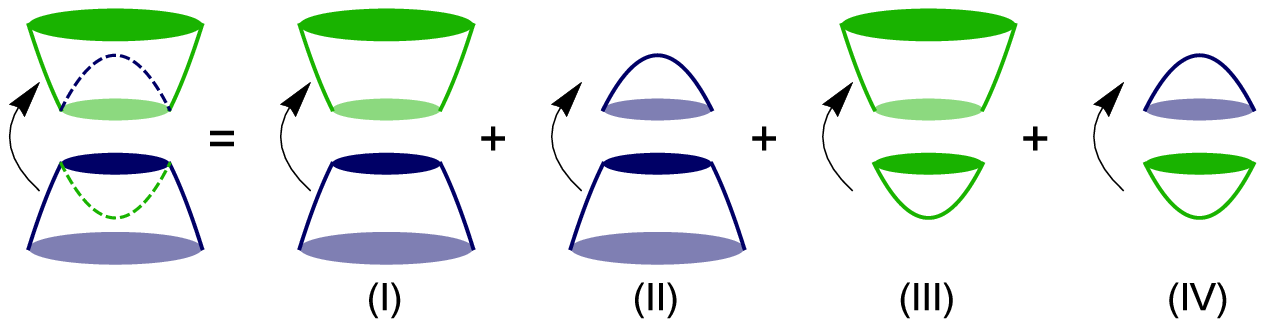}
\par\end{centering}
\centering{}\caption{Schematically decomposing the interband transition between a pair
of inverted bands into four distinct processes in the uninverted band
picture. Processes (I) and (IV) correspond to the interband transitions,
whereas processes (II) and (III) corresponding to the effective intraband
transitions shall greatly enhance the interband coherence and hence
are critical to the strong intrinsic plasmon excitation. \label{FigBI}}
\end{figure} 

The polarization function in Eq. $\left(\ref{polf}\right)$
consists of both the interband and the intraband parts. When the Fermi
level lies in the band gap, at zero temperature, the intraband transition
is forbidden. Thus, the interband correlation will determine the plasmon
excitations. In conventional semiconductors or insulators, the interband
correlation decays rapidly because of small wave function overlap
in Eq. $\left(\ref{overlp}\right)$. This is why the plasmon excitations
usually occur in doped systems, in which the intraband correlation
becomes nonzero and dominates over the interband correlation. 

Here we notice that the band inversion could enhance
the interband correlation and give rise to a strong intrinsic plasmon
excitation. To make the physics more transparent, we divide the
interband transition between the two inverted bands into four different
processes, as shown in Fig.~\ref{FigBI}. It is clear that
the strong overlap of interband transition in the inverted band picture
above corresponds to the effective intraband transition in the uninverted
band picture, labeled by the processes (II) and (III) in Fig.~\ref{FigBI}.  These two processes are responsible for 
the band-inversion enhanced interband transition due to the large wave function overlap [Figs.~\ref{FigBDO}(e)-\ref{FigBDO}(f)].
On the other hand, the joint DOS near the highly degenerate band minima
of this inverted conduction band exhibits an 1D behavior and possesses 
a Van Hove singularity, in which the corresponding DOS diverges, as
shown in Fig. \ref{FigBDO}(c). In addition, there exists a weak discontinuous
point at the conduction band maximum in Fig. \ref{FigBDO}(c). The band-inversion
induced Van Hove singularity has also played a crucial role in unconventional
superconductivity \cite{Goldstein2015SC}, and ferromagnetism \cite{CaoT2015PRL}.
In the following, we show that the large interband correlation
together with the Van Hove singularity is essential to give rise to
strong intrinsic plasmon excitations~\cite{BICM}. 

\begin{figure}
\centering{}\includegraphics[scale=1.05]{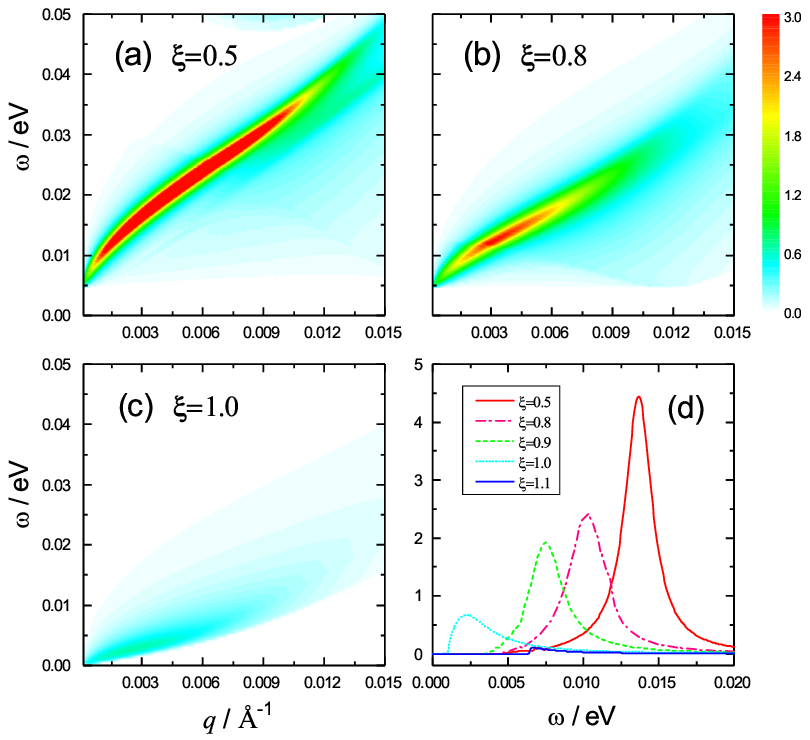}\caption{(color online) The evolution of the plasmon dispersion in the intrinsic
QAH systems of thin films of V- or Cr-doped (Bi,~Sb)$_2$Te$_3$ for $\xi$= 0.5 (a), 0.8 (b), 1.0 (c), respectively. The
color bar denotes intensities of $\mathrm{Loss\left(\mathit{\boldsymbol{q}},\mathit{\omega}\right)}$.
(d) Plots of the energy-loss function at $q=2\times10^{-3}\:\mathbf{\mathrm{\mathring{A}{}^{-1}}}$
for different values of $\xi$. The effective dielectric constant
is $\kappa=5$. Other parameters are the same as those in Fig. \ref{FigBDO}.
\label{Fig3}}
\end{figure}
 
\textit{Plasmon in intrinsic QAH systems.}\textendash To determine
the plasmon dispersion in intrinsic QAH systems of thin films of V- or Cr-doped (Bi,~Sb)$_2$Te$_3$, 
we numerically calculate the energy loss function
$\mathrm{Loss}\left(\mathit{\boldsymbol{q}},\mathit{\omega}\right)$
with several values of $\xi$ and present the corresponding results
in Figs. \ref{Fig3}(a)-\ref{Fig3}(d). When $\xi$ is small, the
system has a clear band inversion structure [see Fig. \ref{FigBDO}(a)].
Meanwhile, there is a strong plasmon excitation in the low frequency
region ($\omega<0.05$ eV), as shown in Fig. \ref{Fig3}(a). The plasmon
dispersion scales approximately linearly with the wave vector $q$.
The energy of this plasmon ranges from several $\mathrm{meV}$ to
tens of $\mathrm{meV}$ and can be effectively tuned by external fields
and other parameters. As $\xi$ increases, the height of the peak
of the Mexican hat decreases. Accordingly, the plasmons becomes weakened
and their dispersion curves become short. The plasmon frequency also
has an obvious red-shift due to the shrink of the gap between the
two inner bands. When the topological phase transition occurs ($\xi$=1),
the band gap closes, and the dispersion curve becomes shortest. Our
numerical calculations also show that, after the topological phase
transition ($\xi$\textgreater{}1), the plasmon becomes almost invisible
due to the weak interband correlation. Thus, the band inversion strongly
modifies the properties of this interband plasmon dispersion. It is
one of the main results in this work. 

\begin{figure}
\begin{centering}
\includegraphics[scale=0.85]{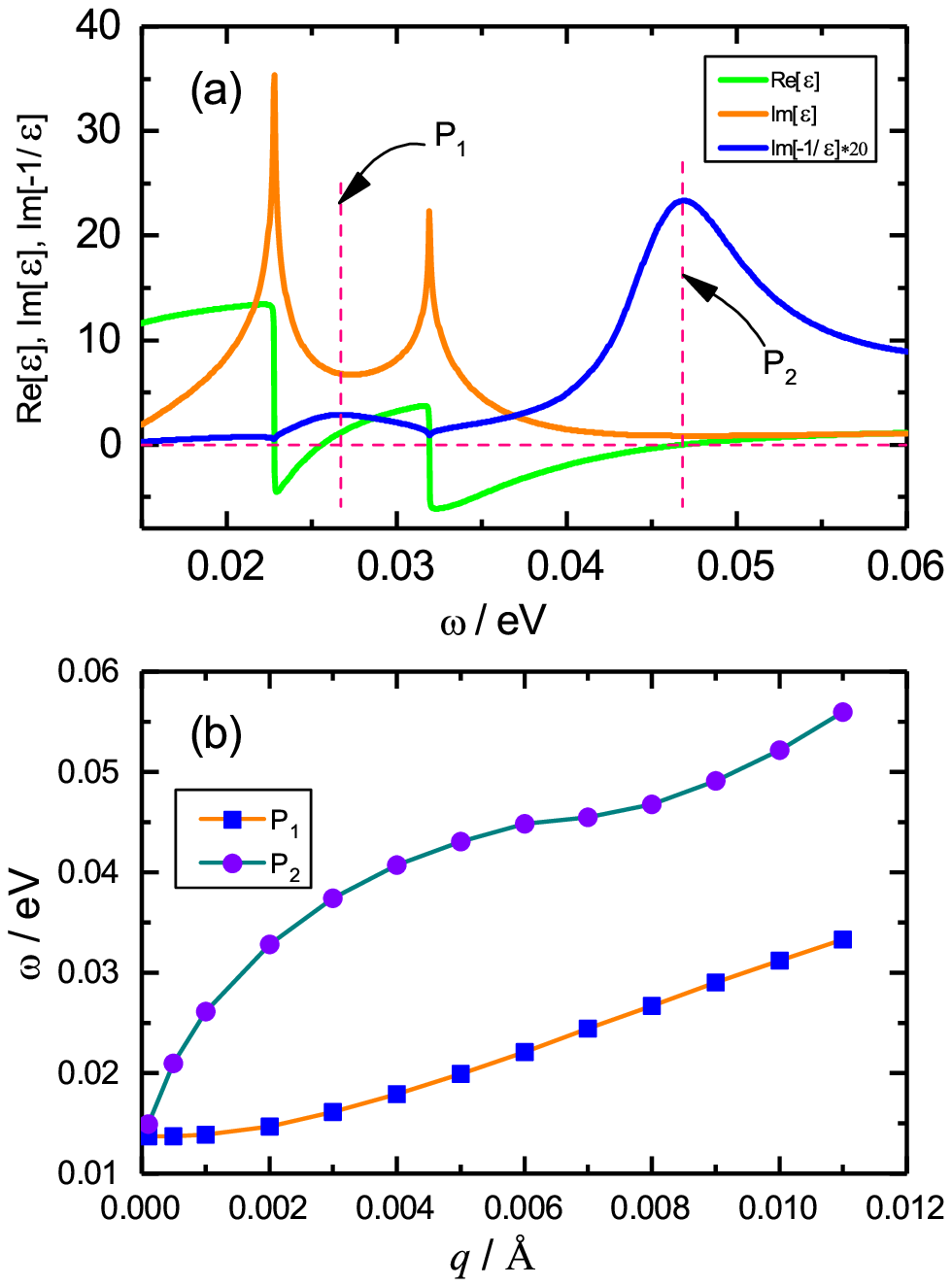}
\par\end{centering}
\centering{}\caption{(color online) The double-peak structure in plasmon dispersion induced by the
electron-hole asymmetric term in the thin films of V- or Cr-doped (Bi,~Sb)$_2$Te$_3$ (a). 
The real part, the  imaginary part of the dielectric function, and the energy-loss function are labeled by the green, yellow and blue lines, respectively. 
(b) shows the energy dispersion of double
plasmons ($\mathrm{P}_{1}$ and $\mathrm{P}_{2}$) determined by the
peaks of the energy loss function. The parameters are $\Delta=-0.02\:\mathrm{eV}$,
$D=28\:\mathrm{eV}\cdot\mathbf{\mathrm{\mathring{A}}}^{2}$, $\kappa=1$,
$\xi=0$, and $q=8\times10^{-3}\:\mathbf{\mathrm{\mathring{A}{}^{-1}}}$.
\label{Fig4}}
\end{figure}

The zeros in the real part of the dynamical dielectric function, $\mathrm{Re[}\varepsilon(q,\omega)]$,
can also be used to determine the plasmon excitations. Fig. \ref{Fig4}(a)
depicts the numerical results of $\mathrm{Re}\left[\varepsilon(q,\omega)\right]$
(green line), $\mathrm{Im}\left[\varepsilon(q,\omega)\right]$ (orange
line) and the energy loss function for $\xi=0$. It is clear that
there exist two peaks (labeled by $\mathrm{P_{1}}$ and $\mathrm{P_{2}}$)
in plasmon dispersion. The plasmon $\mathrm{P_{2}}$ has a higher
energy and a weak damping rate, which is characterized by the imaginary
part of the dielectric function. Compared with $\mathrm{P_{2}}$,
the plasmon $\mathrm{P_{1}}$ has a lower energy and a stronger damping
rate. We also note that a smaller dielectric constant $\kappa$ can
weaken the damping rate and thus enhance plasmon $\mathrm{P_{1}}$.
We plot the dispersions for both plasmons in Fig. \ref{Fig4}(b) and
find that the energy split of this double peak can reach to tens of
$\mathrm{meV}$. Physically, plasmon $\mathrm{P_{2}}$ ($\mathrm{P_{1}}$)
stems from the sum (difference) of processes (II) and (III) in Fig.
\ref{FigBI}. Thus, the double-peak fine structure in the plasmon dispersions
is a direct consequence of the electron-hole asymmetry in Eq. $\left(\ref{H0}\right)$,
which breaks the degeneracy between two interband processes (II) and
(III) in Fig. \ref{FigBI}. It is useful to point out that for
the intrinsic gapped graphene \cite{WangXF2007PRB,Pyatkovskiy2009jpcm},
silicene and other buckled honeycomb lattices \cite{HRChangSilicene,TabertSilicenePRB},
the polarization function within the random phase approximation  satisfies
the relation $\mathrm{Re[}\Pi(q,\omega)]\leq0$ or $\mathrm{Re[}\varepsilon(q,\omega)]\geq1$
\cite{WangXF2007PRB}. As a result, when the Fermi level lies in the
band gap and has a vanishing DOS, these systems hardly develop a plasmon excitation at zero temperature. 

Several remarks are in order here. First, this band inversion enhancement
mechanism for interband plasmons is also applicable to the quantum
spin Hall effect in  HgTe quantum wells \cite{Hasan2010RMP,QiZhangRMP},
and in InAs/GaSb quantum wells \cite{LiuCX2008PRL}. Unlike the effective
model in Eq. $\left(\ref{Ham}\right)$ we study here, the inverted
bands in these systems are usually in the topologically nontrivial
phase. Secondly, in Ref. \cite{Juergens2014PRL}, the dynamical polarization
function and weak plasmons in HgTe quantum wells were calculated.
But the mechanism to enhance and manipulate in-gap plasmon mode was
not discussed. In addition, our present work only reveals the effect
of the band inversion on the longitudinal bulk plasmon excitations.
The case for transverse plasmon excitations will be reported elsewhere.
Note that the plasmon excitations studied here essentially differ
from the edge magnetoplasmon with no magnetic fields observed in the
topological nontrivial phase \cite{mahoney2017EDGE}. 

\textit{Plasmon in lightly doped QAH systems.}\textendash In reality,
the system might be metallic due to defects, self-doping, and charge
transfer from substrates, gating, etc. To gain more insights into
the intrinsic plasmon excitation, we briefly discuss the impact of
finite doping \cite{RBA}. For the lightly doped case, the interband
correlation is still strong enough to rise the novel interband plasmon,
whereas a usual plasmon comes from the intraband transition. On the
other hand, for the heavily doped case, the interband correlation
will be significantly suppressed such that the interband plasmon excitation
disappears. However, the usual intraband plasmon excitation continues to exist. 
\begin{figure}
\includegraphics[scale=1.05]{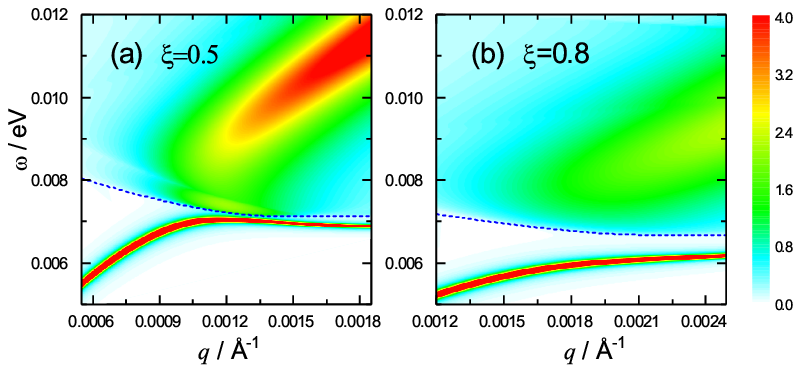}\caption{(color online) The energy dispersion of the plasmons for a lightly
doped QAH system of a thin film of V- or Cr-doped (Bi,~Sb)$_2$Te$_3$ with $\mu=0.004$ eV and $\kappa=5$ for $\xi=0.5$
(a) and 0.8 (b), respectively. The blue dashed lines determined by
minima of the single particle excitation energy, guide to the eye,
denote the edge of electron-hole continuum, whereas the color bar
is for intensities of $\mathrm{Loss\left(\mathit{\boldsymbol{q}},\mathit{\omega}\right)}$.
Other parameters are identical to those in Fig. \ref{FigBDO}. \label{Fig5} }
\end{figure}

Figs.~\ref{Fig5}(a) and \ref{Fig5}(b) depict the energy dispersion of plasmons
for $\mu=4.0$ meV measured from absolute zero. The novel interband
plasmon starts from a relatively large wave vector in Fig. \ref{Fig5}(a)
and becomes weak in Fig. \ref{Fig5}(b) and finally disappears near
the topological phase transition, similar to the intrinsic case in
Fig. \ref{Fig3}. Therefore, the novel interband plasmon still exists
and could signify the topological phase transition in the lightly
doped QAH systems. As shown in Fig. \ref{Fig5}(a), the usual intraband
plasmon starts from a very small wave vector and shows the energy
dispersion $\omega\left(q\right)\propto\sqrt{q}$ for small $q$,
similar to the conventional 2D electron gases \cite{giuliani2005qtel}
and 2D Dirac fermions \cite{grigorenko2012graphene,WangXF2007PRB,Pyatkovskiy2009jpcm,Scholz2012PRB,HRChangSilicene,TabertSilicenePRB}.
After $q$ excesses $1.2\times10^{-3}\:\mathbf{\mathrm{\mathring{A}{}^{-1}}}$,
its dispersion greatly deviates from $\sqrt{q}$. Numerical results
show that the intraband plasmon survives and remains noticeable even
when $\xi>1$. Thus, this intraband plasmon is insensitive to the
topology of energy bands. In addition, the frequency of the intraband
plasmon shows a continuous redshift as increasing $\xi$. 

In summary, it has been shown that the band inversion with the Mexican-hat dispersion indeed
leads to a strong intrinsic plasmon excitation with highly tunable
frequencies in topological materials. The electron-hole asymmetric
term could induce a double-peak fine structure in the plasmon dispersion.
We have further pointed out that this intrinsic plasmon excitation
can be used to characterize the topological phase transition via bulk
measurements. The above physics has been numerically examined in both
the intrinsic and lightly doped thin films of magnetic topological
insulators V- or Cr-doped (Bi,~Sb)$_2$Te$_3$. Therefore, our work paves the way for the potential applications
of topological materials in terahertz and infrared plasmonics. 

A series of recent experimental advances have been made in the detection
of plasmons in 3D topological insulators, in particular, in their
magnetic thin films. We expect that at the current status of experimental
progress, the predicted anomalous plasmons and the double-peak structure
will soon be tested experimentally. 

We are grateful to Hao-Ran Chang, Wen-Yu Shan and Zhi-Ming Yu for
enlightening discussions. F.Z. and Y.Y. acknowledge fundings from
the MOST Project of China (Nos. 2014CB920903 and 2016YFA0300603),
the National Nature Science Foundation of China (Grant Nos. 11574029 and 11734003).
J.Z. and D.X. were supported by AFOSR Grant No. FA9550-14-1-0277.
J.Z. was also supported by the Research Grant Council, University
Grants Committee, Hong Kong under Grant No. 17301116 and C6026-16W.

\bibliographystyle{apsrev4-1}

\begin{thebibliography}{47}
\bibitem [{RBA()}]{RBA}%
  \BibitemOpen
  \href@noop {} {\bibinfo  {journal} {We utilize the rigid-band 
  approximation, in which the energy band structure is assumed not be changed
  by a slightly doping except a shift of the Fermi level. In this work, we
  focus on the n-doped case, namely, the Fermi level crosses the conduction
  bands. The p-doped case can be calculated in a similar way}\ }\BibitemShut {NoStop}%
\end{thebibliography}
%
%
\onecolumngrid

\section*{Supplementary material for ``Tunable Intrinsic Plasmons due to Band Inversion in Topological Materials"}

This supplemental material contains a detailed derivation of Eq. (\ref{polf}) in the main text. We start with the polarization function within the one-loop approximation
at a finite temperature $T$ 
\begin{equation}
\Pi\left(\boldsymbol{q},i\omega_{m}\right)=k_{B}T\sum_{n=-\infty}^{+\infty}\frac{1}{L^{2}}\sum_{\boldsymbol{k}}\mathrm{tr}\left[G\left(\boldsymbol{k},i\Omega_{n}\right)G\left(\boldsymbol{k}^{\prime},i\Omega_{n}+i\omega_{m}\right)\right],
\end{equation}
where $\omega_{m}=2m\pi k_{B}T$ and $\Omega_{n}=\left(2n+1\right)\pi k_{B}T$
are Matsubara frequencies, $L^{2}$ is the area of a 2D system, and
$\boldsymbol{k}^{\prime}=\boldsymbol{k}+\boldsymbol{q}$. Using the
Matsubara Green's function in the eigenstate basis, 
\begin{equation}
G\left(\boldsymbol{k},i\Omega_{n}\right)=\sum_{\lambda}\frac{\left|\lambda\boldsymbol{k}\right\rangle \left\langle \lambda\boldsymbol{k}\right|}{i\Omega_{n}-E_{\lambda\boldsymbol{k}}+\mu},
\end{equation}
with $\mu$ being the chemical potential and $H\left(\boldsymbol{k}\right)\left|\lambda\boldsymbol{k}\right\rangle =E_{\lambda\boldsymbol{k}}\left|\lambda\boldsymbol{k}\right\rangle $,
one gets 
\begin{align}
\Pi\left(\boldsymbol{q},i\omega_{m}\right) & =k_{B}T\sum_{n=-\infty}^{+\infty}\sum_{\lambda,\lambda^{\prime}}\frac{1}{L^{2}}\sum_{\boldsymbol{k}}\sum_{s,\boldsymbol{p}}\left\langle s\boldsymbol{p}\right|\frac{\left|\lambda\boldsymbol{k}\right\rangle \left\langle \lambda\boldsymbol{k}\right|}{i\Omega_{n}-E_{\lambda\boldsymbol{k}}+\mu}\cdot\frac{\left|\lambda^{\prime}\boldsymbol{k}^{\prime}\right\rangle \left\langle \lambda^{\prime}\boldsymbol{k}^{\prime}\right|}{i\Omega_{n}+i\omega_{m}-E_{\lambda^{\prime}\boldsymbol{k}^{\prime}}+\mu}\left|s\boldsymbol{p}\right\rangle .
\end{align}
After making use of the orthogonal relation $\left\langle \lambda\boldsymbol{k}|s\boldsymbol{p}\right\rangle =\delta_{\lambda s}\delta_{\boldsymbol{k}\boldsymbol{p}}$
and summing over $s$ and $\boldsymbol{p}$, one has
\begin{align}
\Pi\left(\boldsymbol{q},i\omega_{m}\right) & =k_{B}T\sum_{n=-\infty}^{+\infty}\sum_{\lambda,\lambda^{\prime}}\frac{1}{L^{2}}\sum_{\boldsymbol{k}}\frac{1}{i\Omega_{n}-E_{\lambda\boldsymbol{k}}+\mu}\cdot\frac{1}{i\Omega_{n}+i\omega_{m}-E_{\lambda^{\prime}\boldsymbol{k}^{\prime}}+\mu}\left|\left\langle \lambda\boldsymbol{k}|\lambda^{\prime}\boldsymbol{k}^{\prime}\right\rangle \right|^{2}.
\end{align}
By use of the summation formula~\cite{giuliani2005qtel}, 
\begin{equation}
k_{B}T\sum_{n=-\infty}^{+\infty}\frac{1}{i\Omega_{n}-\left(x_{1}-\mu\right)}\cdot\frac{1}{i\Omega_{n}+i\omega_{m}-\left(x_{2}-\mu\right)}=\frac{f\left(x_{1}\right)-f\left(x_{2}\right)}{i\omega_{m}+x_{1}-x_{2}},
\end{equation}
performing the summation over $n$ leads to
\begin{align}
\Pi\left(\boldsymbol{q},i\omega_{m}\right) & =\sum_{\lambda,\lambda^{\prime}}\frac{1}{L^{2}}\sum_{\boldsymbol{k}}\frac{f\left(E_{\lambda\boldsymbol{k}}\right)-f\left(E_{\lambda^{\prime}\boldsymbol{k}^{\prime}}\right)}{i\omega_{m}+E_{\lambda\boldsymbol{k}}-E_{\lambda^{\prime}\boldsymbol{k}^{\prime}}}\left|\left\langle \lambda\boldsymbol{k}|\lambda^{\prime}\boldsymbol{k}^{\prime}\right\rangle \right|^{2}.
\end{align}
Carrying out the analytic continuation from Matsubara frequencies
$i\omega_{m}\rightarrow\omega+i\eta$, one obtains the retarded noninteracting
polarization function
\begin{align}
\Pi\left(\boldsymbol{q},\omega\right) & =\sum_{\lambda,\lambda^{\prime}}\frac{1}{L^{2}}\sum_{\boldsymbol{k}}\frac{f\left(E_{\lambda\boldsymbol{k}}\right)-f\left(E_{\lambda^{\prime}\boldsymbol{k}^{\prime}}\right)}{\omega+E_{\lambda\boldsymbol{k}}-E_{\lambda^{\prime}\boldsymbol{k}^{\prime}}+i \eta}\left|\left\langle \lambda\boldsymbol{k}|\lambda^{\prime}\boldsymbol{k}^{\prime}\right\rangle \right|^{2}.
\end{align}
Replacing $\left(1/L^{2}\right)\sum_{\boldsymbol{k}}$ with $\int\frac{d^{2}\boldsymbol{k}}{\left(2\pi\right)^{2}}$,
one immediately reaches 
\begin{align}
\Pi\left(\boldsymbol{q},\omega\right) & =\sum_{\lambda,\lambda^{\prime}}\int\frac{d^{2}\boldsymbol{k}}{\left(2\pi\right)^{2}}\frac{\left[f\left(E_{\lambda\boldsymbol{k}}\right)-f\left(E_{\lambda^{\prime}\boldsymbol{k}^{\prime}}\right)\right]}{\omega+E_{\lambda\boldsymbol{k}}-E_{\lambda^{\prime}\boldsymbol{k}^{\prime}}+i\eta}\left|\left\langle \lambda\boldsymbol{k}|\lambda^{\prime}\boldsymbol{k}^{\prime}\right\rangle \right|^{2},
\end{align}
which is the noninteracting polarization function in Eq. $\left(4\right)$
in the main text. 
\end{document}